\documentclass[a4paper,11pt]{article}
\pdfoutput=1 

\usepackage{jinstpub} 

\usepackage[english]{babel}
\usepackage{amsmath}        
\usepackage{wasysym}        
\usepackage{gensymb}        
\usepackage[dvipsnames]{xcolor}
\usepackage{graphicx}       
\usepackage{dcolumn}        
\usepackage{subcaption}     
\usepackage{xspace}         
\usepackage{nicefrac}       
\usepackage{units}          
\usepackage{hyperref}       
\usepackage{multirow}       
\usepackage{lineno}         
\usepackage{ulem}           

\usepackage{tikz}               
\usetikzlibrary{arrows.meta}    

\hypersetup{
    pdfnewwindow=true,  
    colorlinks=true,    
    linkcolor=blue,     
    citecolor=blue,     
    filecolor=blue,     
    urlcolor=blue       
}

\graphicspath{{img/}}
\newcommand{\umolpercent}{\ensuremath{\%_\mathrm{mol}}\xspace}
\newcommand{\upbar}{\ensuremath{\mathrm{bar}}\xspace}
\newcommand{\upmbar}{\ensuremath{\mathrm{m\upbar}}\xspace}
\newcommand{\uEfield}{\ensuremath{\nicefrac{\uVolt}{\mathrm{cm}}}\xspace}
\newcommand{\uKelvin}{\ensuremath{\mathrm{K}}\xspace}
\newcommand{\uCelsius}{\ensuremath{\degree\mathrm{C}}\xspace}
\newcommand{\uliter}{\ensuremath{\mathrm{\ell}}\xspace}
\newcommand{\uCentimeter}{\ensuremath{\mathrm{cm}}\xspace}
\newcommand{\uMillimeter}{\ensuremath{\mathrm{mm}}\xspace}
\newcommand{\uMicrometer}{\ensuremath{\mu\mathrm{m}}\xspace}
\newcommand{\uVolt}{\ensuremath{\mathrm{V}}\xspace}
\newcommand{\ueV}{\ensuremath{\mathrm{eV}}\xspace}
\newcommand{\uNanoSecond}{\ensuremath{\mathrm{ns}}\xspace}

\newcommand{\uETp}{\ensuremath{\nicefrac{\uVolt}{\uCentimeter} \, \nicefrac{\uKelvin}{\upmbar}}\xspace}

\newcommand{\uEpbar}{\ensuremath{\nicefrac{\uVolt}{\uCentimeter} \, \nicefrac{1}{\upbar}}\xspace}

\newcommand{\uvd}{\ensuremath{\nicefrac{\uMicrometer}{\mathrm{ns}}}\xspace}
\newcommand{\ukiloVolt}{\ensuremath{\mathrm{kV}}\xspace}
\newcommand{\uMeV}{\ensuremath{\mathrm{M}\ueV}\xspace}
\newcommand{\uMeVc}{\ensuremath{\uMeV/c}\xspace}

\newcommand{\ETp}{\ensuremath{\nicefrac{\mathrm{ET}}{\mathrm{p}}}\xspace}
\newcommand{\Ep}{\ensuremath{\nicefrac{\mathrm{E}}{\mathrm{p}}}\xspace}

\newcommand{\ESTP}{\ensuremath{\mathrm{E}_{\mathrm{STP}}}\xspace}
\newcommand{\vd}{\ensuremath{v_\mathrm{d}}\xspace}
\newcommand{\estar}{\ensuremath{\mathrm{E}^\ast}\xspace}

\newcommand{\hptpc}{\mbox{\textrm{HPTPC}}\xspace}
\newcommand{\pten}{\mbox{\textrm{P10}}\xspace}

\newcommand{\hpgmc}{\mbox{\textrm{HPGMC}}\xspace}
\newcommand{\dune}{\mbox{\textrm{DUNE}}\xspace}
\newcommand{\magboltz}{\mbox{\texttt{\textsc{MagBoltz}}}\xspace}
\newcommand{\agros}{\mbox{\texttt{\textsc{Agros2D}}}\xspace}

\newcommand{\strontium}{\ensuremath{{}^{90}\mathrm{Sr}}\xspace}
\newcommand{\yttrium}{\ensuremath{{}^{90}\mathrm{Y}}\xspace}
\newcommand{\iron}{\ensuremath{{}^{55}\mathrm{Fe}}\xspace}
\newcommand{\vdmax}{\ensuremath{\vd^\mathrm{max}}\xspace}

\newcommand{\methane}{\ensuremath{\mathrm{CH}_4}\xspace}

\begin{document}

\title{A Gas Monitoring Chamber for High Pressure Applications}

\author[a,1]{Philip Hamacher-Baumann\note{Corresponding author.}}
\emailAdd{hamacher.baumann@physik.rwth-aachen.de}
\author[a]{Stefan Roth}
\author[a]{Thomas Radermacher}
\author[a]{Nick Thamm}
\affiliation[a]{III. Physikalisches Institut, RWTH Aachen University, 52056 Aachen, Germany}

\abstract{
    Time Projection Chambers (TPCs) operated at high pressure have become a topic of interest for future long baseline neutrino experiments.
    Pressurized gas retains the low momentum threshold for particle detection of atmospheric TPCs, but offers a larger target mass for neutrino interactions at the same volume.
    Operation at high pressure poses several new challenges in safety aspects regarding overpressure and high voltage safety.
    The presented  High Pressure Gas Monitoring Chamber (\hpgmc) can be used to study the suitability of various drift gas mixtures up to \unit[10]{bar} and a maximum field of $\sim\unit[3000]{\nicefrac{\mathrm{V}}{\mathrm{cm}}}$.
    A flexible construction makes it possible to exchange parts of the inner detector and to test new technologies.
    In this work, the construction of a \hpgmc and its commissioning using the P10 gas mixture (\unit[90]{\%} Ar + \unit[10]{\%} \methane) are presented.    
}

\arxivnumber{2005.03636}

\maketitle

\section{Introduction}\label{sec:intro}
Time Projection Chambers (TPCs) are gaseous detectors that have originally been conceived for tracking moderate numbers of particles in clean lepton interactions at colliders~\cite{Hilke:2010zz}.
Since then, the TPC technology has evolved and is presently employed from rare event searches~\cite{Gonzalez:2018rar} to ultra-high multiplicities nuclear collision experiments~\cite{Alme:2010ke}.
The gas is interchangeable and can be tailored to the given task without modification of the main detector parts.
Since the active medium is of very low density, a low momentum, below \unit[200]{\uMeVc} for protons, threshold for track reconstruction can be achieved.
Hence, TPCs have become a topic of interest for experiments at the few GeV energy scale, where the range of particles limits resolution, e.g. for separation of tracks at an interaction vertex.

Currently, the modelling of nuclear effects in final state interactions are the limiting systematic uncertainties of long baseline neutrino experiments~\cite{Alvarez-Ruso:2017oui}.
It can be improved by new data in the low momentum region~\cite{nd280_upgrade_tdr}.
To achieve a higher target mass, while retaining a low momentum threshold, High Pressure Time Projections Chambers (\hptpc) have been proposed for future experiments.
The Deep Underground Neutrino Experiment's (\dune) near detector complex foresees to build a \unit[10]{\upbar} \hptpc to constrain interaction uncertainties in combination with the liquid argon filled far detectors~\cite{Abi:2020wmh,AbedAbud:2021hpb}.
This work presents the construction and commissioning of a small-scale gaseous detector aimed at studies of electron drift parameters at high pressures.

\section{Gas Monitoring Chambers}\label{sec:gmc}
The characteristics of the TPC's drift gas are a necessary input for the initial design and for reconstruction in a running detector.
Any change of drift gas properties will affect performance.
Some TPCs have life times in excess of \unit[10]{years} under constantly changing environmental conditions (T, p) and gas quality.
Hence, continuous calibration of electron drift properties is needed to mitigate mid to long term deviations from design values.
Gas Monitoring Chamber (GMC) detectors have been used for that purpose by many experiments, including L3~\cite{GOETTLICHER_1991}, OPAL~\cite{Huk:1988}, CMS~\cite{Reithler:2018} and T2K~\cite{Abgrall:2010hi}.

\subsection{Working Principle}\label{ssec:gmc_principle}
\begin{figure}[!ht]
    \begin{subfigure}{.49\linewidth}
        \centering
        \def\svgwidth{.8\columnwidth} 
        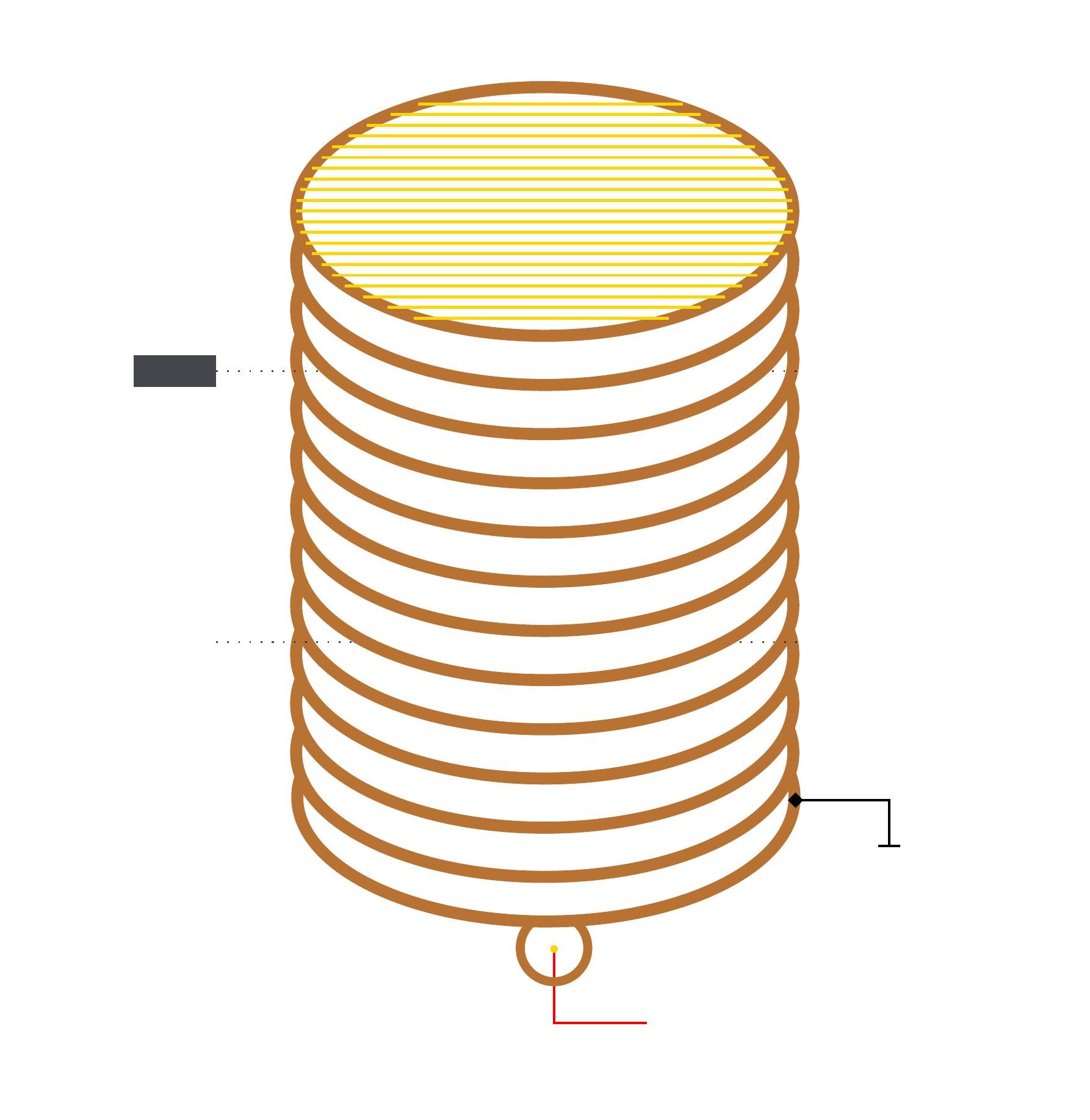
        \caption{}
        \label{sfig:gmc_sketch}
    \end{subfigure}
    \hfill%
    \begin{subfigure}{.45\linewidth}
        \centering
        \includegraphics[width=.9\columnwidth]{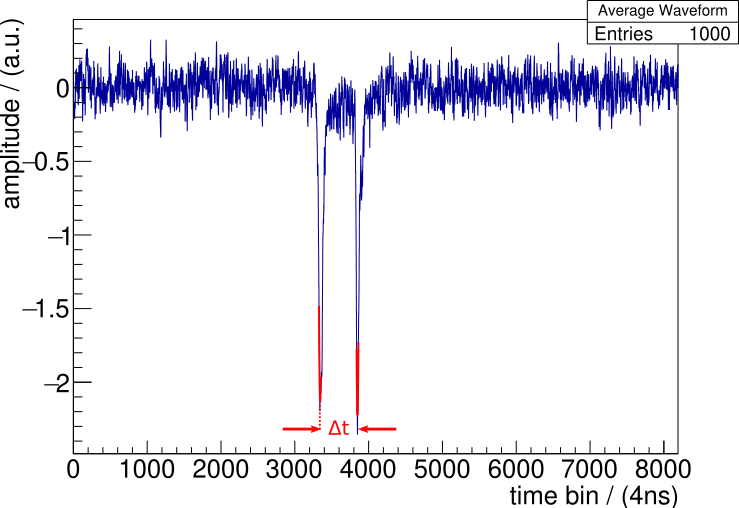}
        \caption{}
        \label{sfig:gmc_signal}
    \end{subfigure}
    \caption{
        (a) Sketch of the \hpgmc with radioactive sources for drift velocity measurement.
        The field cage has a length of $\sim\unit[97]{\uMillimeter}$, an inner diameter of $\sim\unit[77]{\uMillimeter}$ and the sources are separated by $\Delta z = \unit[51.95]{\uMillimeter}$.
        Applied drift fields can mirror that of a host TPC or systematically scan through different fields to create a drift velocity curve.
        (b) Average over 1000 waveforms recorded at the anode taken in \pten (\unit[90]{\%} Ar + \unit[10]{\%} \methane) at \unit[1.5]{\upbar} and with the \hpgmc.
        The time difference of the two peaks is the drift time between the near and far source's position.
    }
    \label{fig:gmc_schema}
\end{figure}

A GMC is a miniature TPC where ionization tracks are created at known positions by radioactive sources (\strontium).
Emitted $\beta$-electrons can traverse the drift volume inside a field cage, leaving a track of liberated electrons and ions, and exit into a scintillating fibre, see figure~\ref{sfig:gmc_sketch}.
The light signal created in the scintillating fibres is used to start a fixed time window readout of signals from avalanche-multiplied electrons recorded at the GMC's anode.
Figure~\ref{sfig:gmc_signal} shows the average of 1000 recorded waveforms without distinguishing between the two (near and far) source positions.
The drift velocity \vd is then calculated from the time difference $\Delta t$ between the two signal peaks and the known distance $\Delta z$ between near and far source locations:
\begin{equation}
    \vd = \frac{\Delta z}{\Delta t}
    \label{eq:vd_definition}
\end{equation}

In this scheme, the drift velocity is effectively measured between the source positions in a central drift region of the GMC.
This reduces or removes major systematic effects from the measurement.
The central field cage region has a more homogeneous drift field than at the edges of the chambers, close to the field cage walls.
In addition, drift velocity deviations from strong fringe fields close to the anode cancel out.
Furthermore, constant-time response delays of the readout, i.e. from cables and trigger electronics, are removed, because only time differences enter in~[Eq.~\ref{eq:vd_definition}].

GMCs have also been equipped with mono-energetic X-ray sources to monitor gas gain.
By continuous measurement of the spectrum of \iron, deviations in the gas gain can be monitored and calibrated for or corrected~\cite{Abgrall:2010hi}.
Owing to the small gas volumes of GMCs of $\unit{\sim 1}{\uliter}$ to $\unit{5}{\uliter}$, gas can quickly be replaced without loosing much time or gas for flushing.
This feature has been exploited to monitor multiple sub-detectors in sequence~\cite{Reithler:2018}.

\subsection{Continuous Calibrations}\label{ssec:density_corr}
Drifting electrons are accelerated and deflected by electric and magnetic fields between collisions with the gas atoms or molecules.
The rate at which these collisions take place depends on the gas density and follows pressure and temperature fluctuations.
The drift velocity then only depends on the ratio of the electric field and gas density~\cite{BlumRolandi:2008}.
It is routinely corrected for by evaluating gas properties at the scaled electric field \estar:
\begin{equation}
    \vd(\estar) = \vd\left(\frac{ET}{p}\right).
    \label{eq:vd_etp}
\end{equation}
The quantity \estar is often referred to as the reduced electric field.

TPCs are often not well controlled in pressure or temperature, but follow environmental fluctuations~\cite{Abgrall:2010hi}.
As a result, the drift properties of the used gas mixture vary over time, and have to be accounted for when reconstructing particle tracks or combining data sets over long running periods.
One strategy used is to operate a GMC in parallel to a main detector TPC with gas supplied by the detector's return line and identical electric field~\cite{Abgrall:2010hi}.
Temperature and pressure of the gas in the TPC and GMC are recorded.
Both data sets can be calibrated towards a common $T_0$ and $p_0$ to make the data comparable over the runtime by correcting changes in gas density.
Changes through varying gas quality and ageing to some extent are picked up and calibrated for by this strategy.
For the T2K near detector ND280, the gas gain fluctuations are reduced to \unit[1]{\%} compared to $\sim\unit[10]{\%}$ without calibration~\cite{Abgrall:2010hi}.
The remaining fluctuations are thought to be caused by other influences on the gas gain, such as the mixing stability~\cite{nd280_upgrade_tdr}.

\section{High Pressure Operation}\label{sec:HPop}
The operation principle of an atmospheric GMC can be transferred almost identically to a High Pressure Gas Monitoring Chamber (\hpgmc).
The target pressure is taken to be \unit[10]{\upbar}, following the design of the DUNE near detector~\cite{AbedAbud:2021hpb}.

Many commonly used quenching gases are flammable, adding another source of danger to high overpressure.
To ensure safe operation, the critical parts are sourced from industry with certified and verified quality.
Pressure-holding parts are manufactured according to DIN standard, which restricts the vessel's capacity to a maximum of \unit[50]{\upbar$\ell$}, or \unit[5]{$\ell$} at \unit[10]{\upbar}~\cite{DIN_EN_13445_1}.
A small inner volume is also desirable to provide a frequent exchange of the gas inside and thus low latency detection of changes in the gas mixture composition or quality (Section~\ref{ssec:gmc_principle}).

\subsection{Range of Electrons in High Pressure Gas}\label{ssec:range}
\begin{figure}[htbp]
    \begin{subfigure}{.65\linewidth}
        \centering
        \includegraphics[width=\columnwidth]{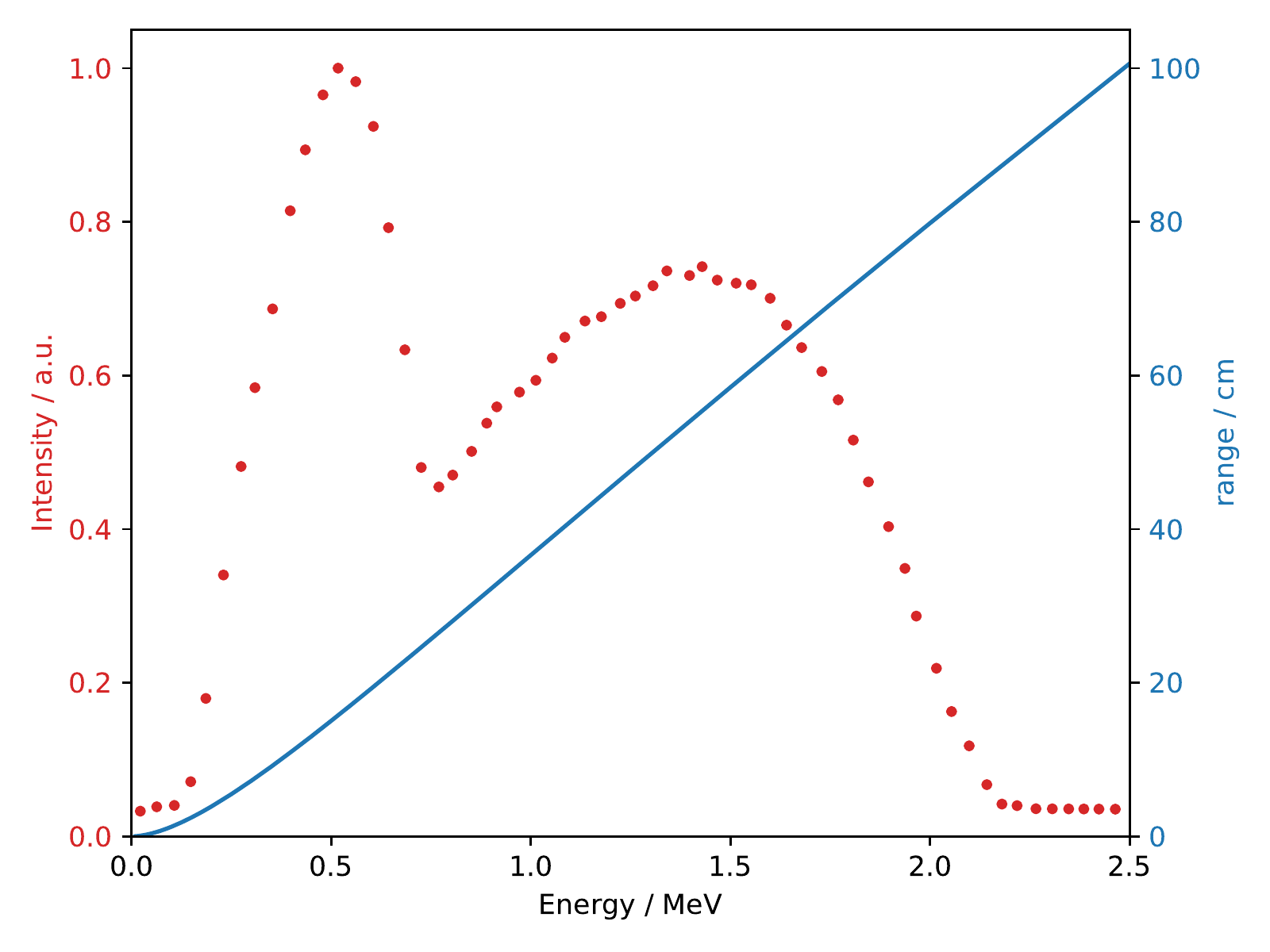} 
        \caption{}
        \label{sfig:hp_range}%
    \end{subfigure}
    \hfill%
    \begin{subfigure}{.4\linewidth}
        \centering
        \def\svgwidth{\columnwidth}
        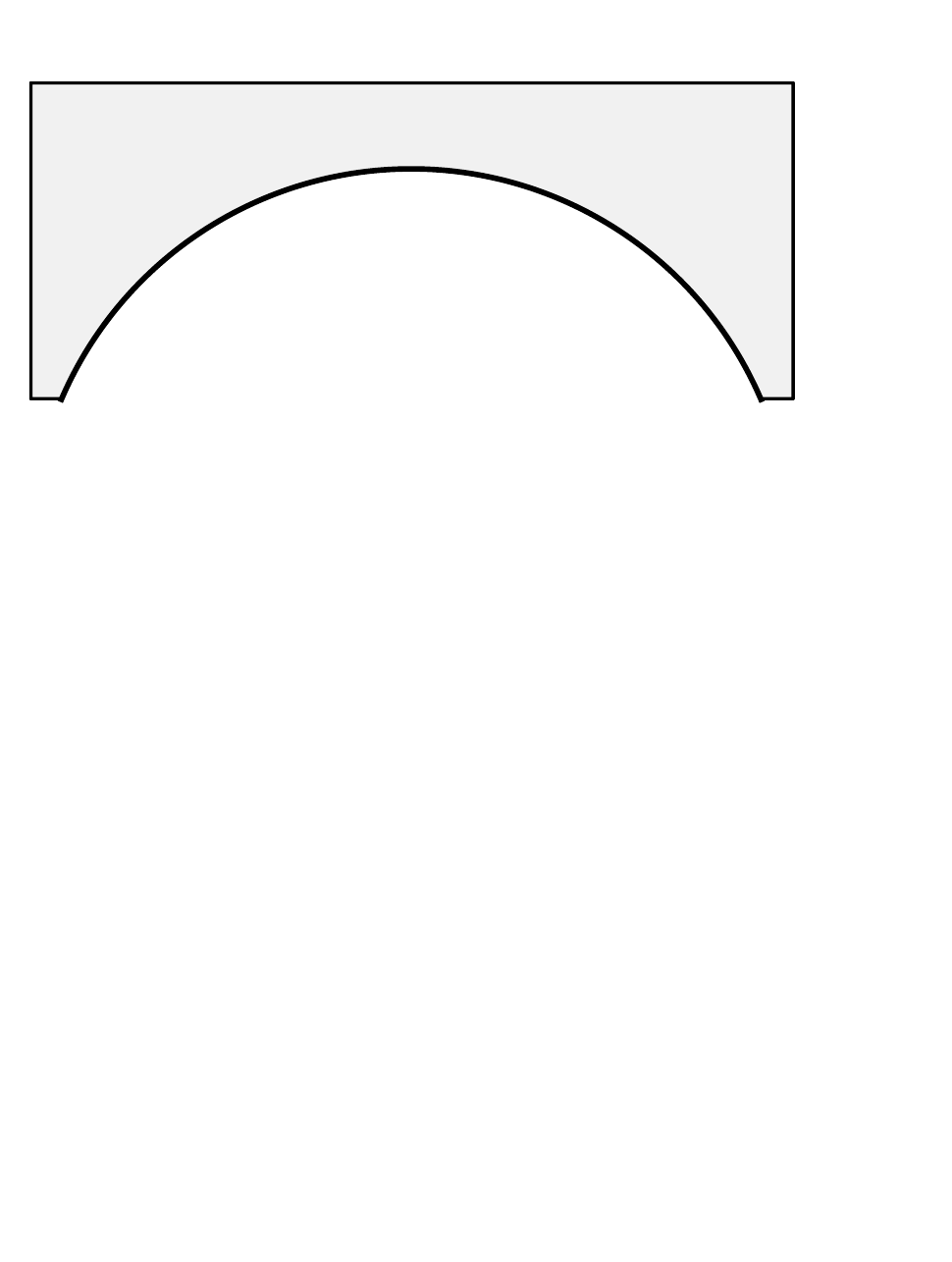
        \caption{}
        \label{sfig:xsec}
    \end{subfigure}
    \caption{%
        (a) Range in \unit[10]{\upbar} Ar (blue) and spectrum (red, from~\cite{Hansen:1983}) of electrons emitted by the \strontium decay chain.
        The two peaks correspond to the \strontium (left) and \yttrium (right) decays in the source.
        To retain reasonable rates, the traversed distance in the gas should not be larger than \unit[10]{\uCentimeter}.
        Values computed using CSDA range listed in NIST's ESTAR database~\cite{STAR:2017db}.
        (b) Cross section view of the field cage where the $\beta$-electrons cross the drift volume.
        The collimation bores and opposing fibres are exposed to the drift space between the field cage rings.
        Both source positions are geometrically identical.
    }
    \label{fig:electron_path}
\end{figure}

At high pressures, the electrons from the \strontium sources are strongly attenuated in range and scattered due to the increased gas density and more often prevented from reaching the start-trigger fibres than at atmospheric pressure.
A considerable fraction of the $\beta$-electrons are produced with an energy around \unit[500]{keV}~\cite{Hansen:1983}, that have an approximate range of \unit[10]{\uCentimeter} in \unit[10]{\upbar} argon~\cite{STAR:2017db}, see figure~\ref{sfig:hp_range}.
It is necessary, that $\beta$-electrons can penetrate the pressurized gas to reach the start-trigger fibres and then still carry enough energy to produce scintillation light.
Therefore, the distance which the $\beta$-electrons have to cross from source to fibre was chosen to be smaller than \unit[10]{\uCentimeter}.
The electrons enter the field cage in a collimated beam that is further widened by multiple scattering.
Opposing the \strontium sources, and their collimation, are $\unit[8 \times 1]{\mathrm{mm}^{2}}$ start-trigger fibres aligned such that only tracks are selected, that have not been deflected more than one fibre height (\unit[1]{mm}) along the drift direction (compare figure~\ref{sfig:xsec}).

A more fundamental difference to an atmospheric GMC comes from the fact that the sources have to be placed inside the pressurized volume, as the $\beta$-electrons can not penetrate the pressure vessel -- a \unit[2.5]{MeV} electron has a range of only $\sim\unit[2]{mm}$ in stainless steel~\cite{STAR:2017db}, which is less than the vessel wall thickness (section~\ref{sec:hpgmc}).
The sources are classified as ISO/12/C64344~\cite{ISO2919:2012}, which certifies save operation up to \unit[70]{\upbar} overpressure.

\subsection{High Voltage Safe Operation}\label{ssec:hv_safety}
The drift velocity is a function of the reduced field \estar [Eq.~\ref{eq:vd_etp}].
An increase in pressure will expand drift velocity curves along the electric field axis.
This effect can be cancelled by a proportional increase of the drift field $E$ to keep \estar constant.
Typical drift fields are of the order of a few \unit[100]{\uEpbar} (\unit[30]{\uETp}) -- to reach the corresponding reduced fields at \unit[10]{\upbar}, the \hpgmc is designed for a maximum cathode voltage of \unit[30]{kV} with a $\unit[\sim10]{\uCentimeter}$ long field cage.

One concern with high voltages are electric breakdowns, that could damage parts of the detector or readout electronics.
The breakdown in a gas can be calculated by Paschen's Law
\begin{align}
    V_\mathrm{b}(pd) =%
    \frac{\mathrm{B}pd}%
    {\ln\left(\mathrm{A}pd\right) - \ln\left[\ln\left(1+\gamma_{\mathrm{se}}^{-1}\right)\right]}\,,
    \label{eq:paschen_law}
\end{align}
that predicts the breakdown voltage $V_\mathrm{b}$ between two idealized parallel surfaces separated by a gas at pressure $p$, the separation distance $d$, gas-specific parameters $\mathrm{A}, \mathrm{B}$ and a material constant $\gamma_{\mathrm{se}}$~\cite{Paschen:1889, liebermann:2005}.
The parameters $\mathrm{A,B}$ are experimentally found to be constant over a limited range of \Ep~\cite{liebermann:2005}.
For scans over larger ranges, $\mathrm{A,B}$ can not be assumed to be constant, but can be calculated from Townsend simulations of pure gasses, e.g. with \magboltz~\cite{Biagi:1999magb,liebermann:2005,HamacherBaumann:2017msc}.

The secondary electron emission coefficient $\gamma_{\mathrm{se}}$ describes how many electrons are released from ions impacting on the cathode and reach the anode.
It depends on the involved electrode materials and is highest for alkali metals~\cite{liebermann:2005}.
Its influence on the resulting $V_\mathrm{b}$ is negligible for high voltages and pressures, where
\begin{align}
    \label{eq:hp_breakfown_evolution}
    V_\mathrm{b} \propto pd
\end{align}

\begin{figure}[!ht]
    \begin{subfigure}{.49\linewidth}
        \centering
        \includegraphics[width=\linewidth]{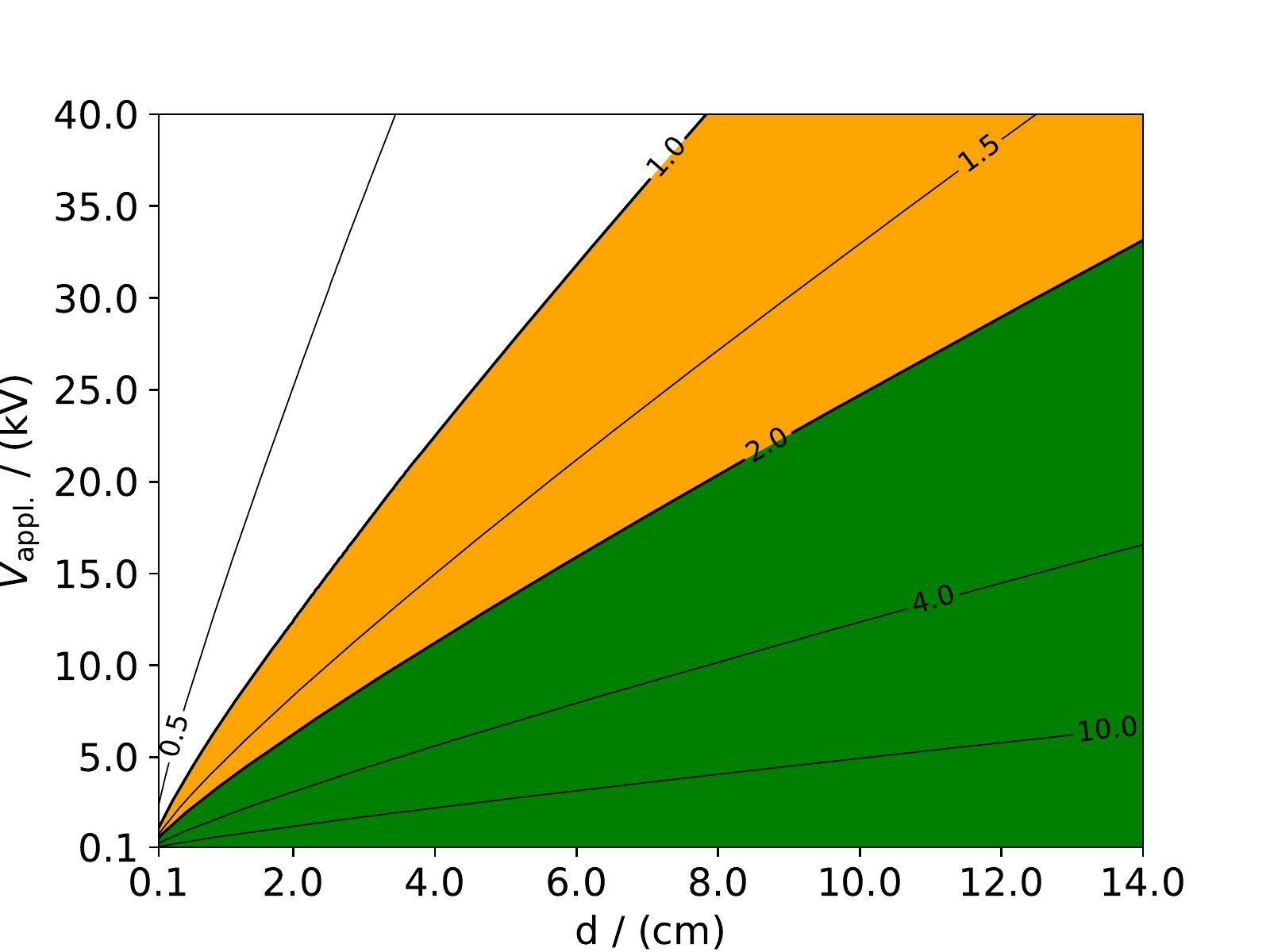}
        \caption{Ar at \unit[1]{\upbar}}
        \label{sfig:breakdown_1bar}
    \end{subfigure}
    \hfill%
    \begin{subfigure}{.49\linewidth}
        \centering
        \includegraphics[width=\linewidth]{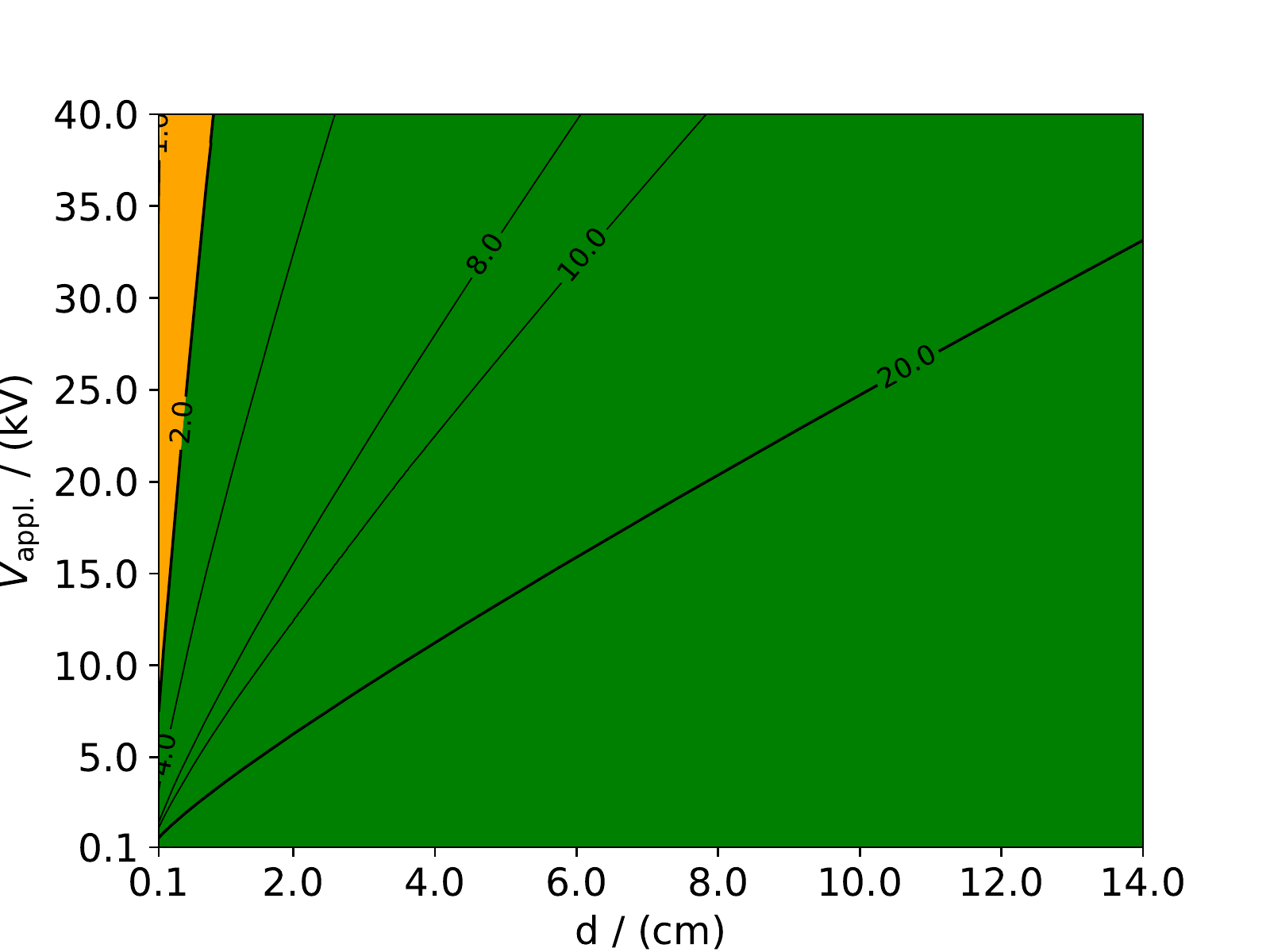}
        \caption{Ar at \unit[10]{\upbar}}
        \label{sfig:breakdown_10bar}
    \end{subfigure}
    \caption{
        Contours of constant ratio of breakdown voltage to voltage applied over a gap $d$ in (a) \unit[1]{\upbar} and (b) \unit[10]{\upbar} argon (figure from~\cite{HamacherBaumann:2017msc}).
        Distances are chosen such that ${V_\mathrm{b}}/{V_\mathrm{applied}}$ stays above 2 (green region).
        The orange region ($1\leq{V_\mathrm{b}}/{V_\mathrm{applied}}\leq2$) is avoided as a safety buffer.
        Isolation distances for high pressure argon are significantly smaller.
    }
    \label{fig:breakdown_factor}
\end{figure}

The fraction ${V_\mathrm{b}}/{V_\mathrm{applied}}$ is used to determine a safe operation limit by choosing insulation distances such that 
\begin{equation}
    V_\mathrm{b}(pd)\overset{!}{>}2\cdot V_\mathrm{applied}\\%
    ,\quad\mathrm{with }\,p=\unit[1]{\upbar}.
    \label{eq:safety_zone}
\end{equation}
In figure~\ref{fig:breakdown_factor}, contours of constant fractions between applied and breakdown voltage are calculated by evaluating [Eq.~\ref{eq:safety_zone}]. 
At the maximum design voltage of \unit[30]{kV} for the \hpgmc, the clearance between the cathode on high voltage and the pressure vessel on ground potential would have to be larger than \unit[12]{\uCentimeter}.
This would introduce too much dead space to comply with a limit of $\unit[5]{\ell}$ for the inner volume (section~\ref{sec:HPop}).
Therefore, the vessel walls of the \hpgmc are lined with \unit[4]{mm} PFA, a PTFE derivative with a high dielectric strength of \unit[80]{kV/mm}~\cite{DuPont:2017PFA} and low outgassing~\cite{NASA:2017otg}.
Argon at \unit[10]{\upbar} pressure requires about $\nicefrac{1}{10}$ of the insulation distance of atmospheric argon~[Eq.~\ref{eq:hp_breakfown_evolution}].
Initially, the requirement from [Eq.~\ref{eq:safety_zone}] was arbitrarily chosen.
To date, we have operated the \hpgmc often at its design maximum, and never experienced breakdown of the gaseous insulators.

\begin{figure}[!htbp]
    \centering
    \includegraphics[trim={0 0 0 6.5cm}, clip, width=.8\columnwidth]{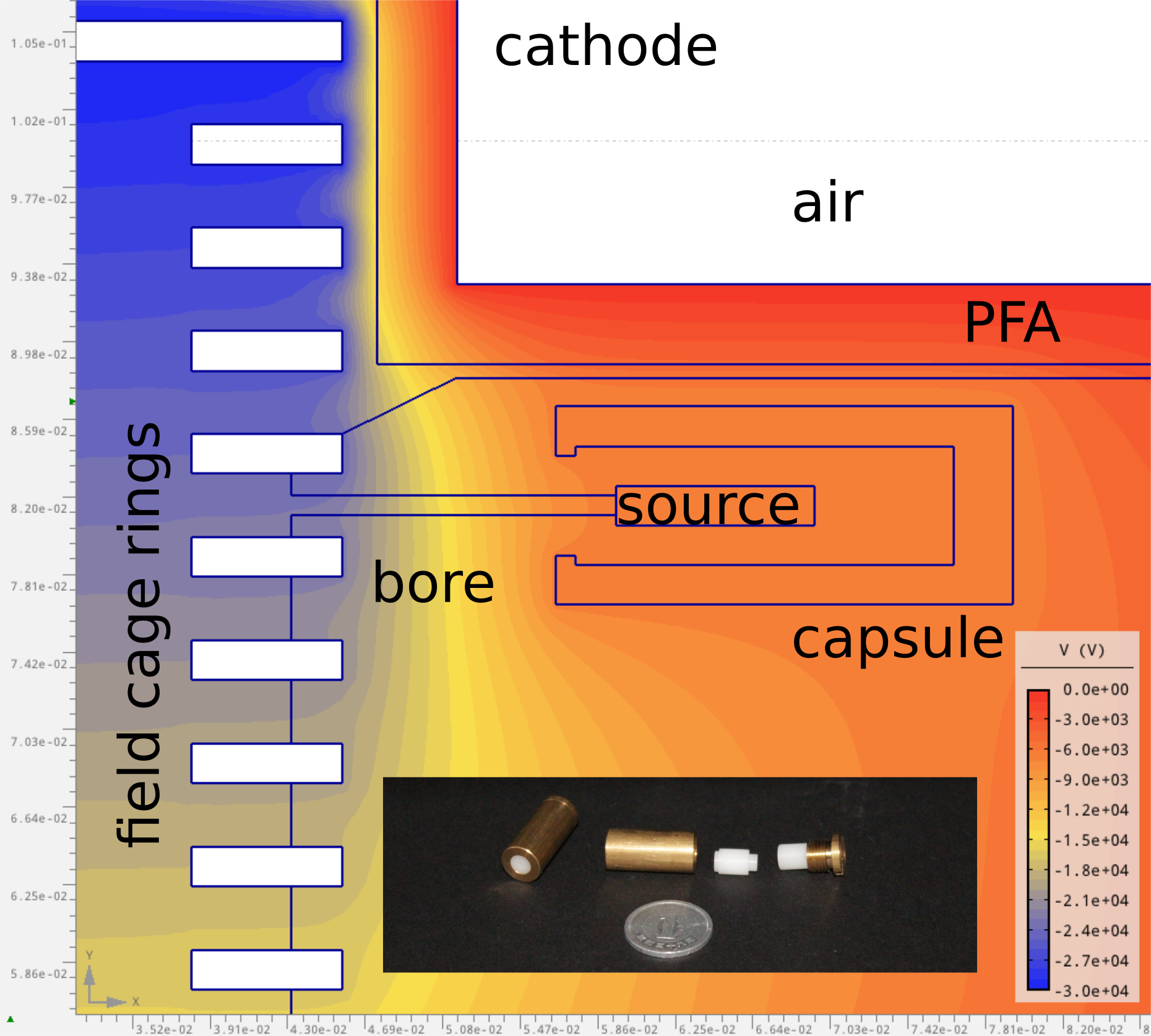}%
    \caption{
        Electrostatic simulation for the far source position, done with \agros (figure from~\cite{HamacherBaumann:2017msc}).
        The cathode has been fixed at the maximum voltage of \unit[30]{\ukiloVolt}, field rings at their corresponding degraded voltage and the pressure vessel outside the PFA cladding is grounded.
        Source and capsule are metallic, but left electrically floating within the POM holding structure.
        The calculated breakdown field inside the bore is always more than twice of the actual field.
        The inset shows the actual sources holders with a \unit[1]{Yen} coin for scale (\unit[20]{mm} diameter).
        The holding structures are held in place by the rigid field cage.
    }
    \label{fig:sources_hv_safety}
\end{figure}

Special care has to be taken for the location of the radioactive sources.
To retain a reasonable signal rate, the sources have to be as close as possible to the drift volume.
This exposes them to voltages up to \unit[30]{kV} for the far source, which sits close to the cathode plane.
Electrostatic simulations with \agros~\cite{karban:2013agros} have been performed for the far source position to determine the minimum required distance from [Eq.~\ref{eq:safety_zone}].
The $\unit[2]{mm} \times \unit[10]{mm}$ sources are installed in brass capsules, which then are placed close to the field cage inside a solid holder made from POM\footnote{Polyoxymethylene, often known by the trade name Delrin\textsuperscript{\textregistered}} (figure~\ref{fig:sources_hv_safety} and section~\ref{sec:hpgmc}).
The near position is always exposed to lower voltages, but is placed at the same distance from the field cage as the far source.

\section{The High Pressure Gas Monitoring Chamber: \hpgmc}\label{sec:hpgmc}
\begin{figure*}[!htbp]
    \centering
    \includegraphics[width=.9\linewidth]{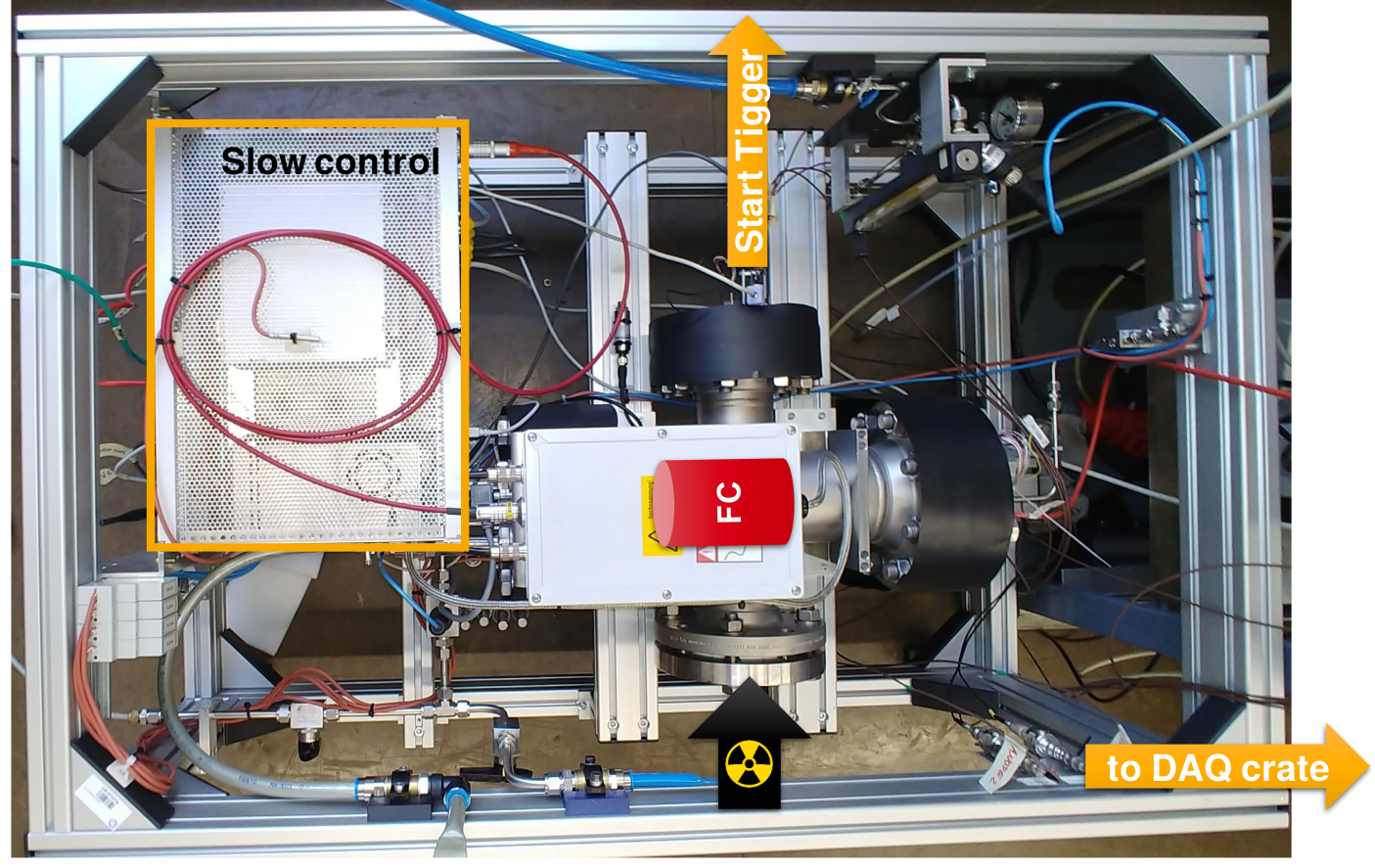}
    \caption{
        \hpgmc mounted in a euro-pallet sized crate.
        The flanged cross piping pressure vessel is located in the center under a grey HV distribution box.
        An integrated \unit[19]{"} slow control rack contains cathode HV and sensor readout.
        The field cage (FC) is placed in the centre of the cross with radioactive sources and trigger electronics installed in the side arms on opposing sides.
        Not shown is the DAQ crate, that also holds the HV module for the anode voltage.
    }
    \label{fig:hpgmc_crate}
\end{figure*}

\begin{figure}
    \centering
    \begin{tikzpicture}
        \node[anchor=south west,inner sep=0] (image) at (0,0) {%
            \includegraphics[trim={0 13cm 0 10cm}, clip,width=.75\columnwidth]{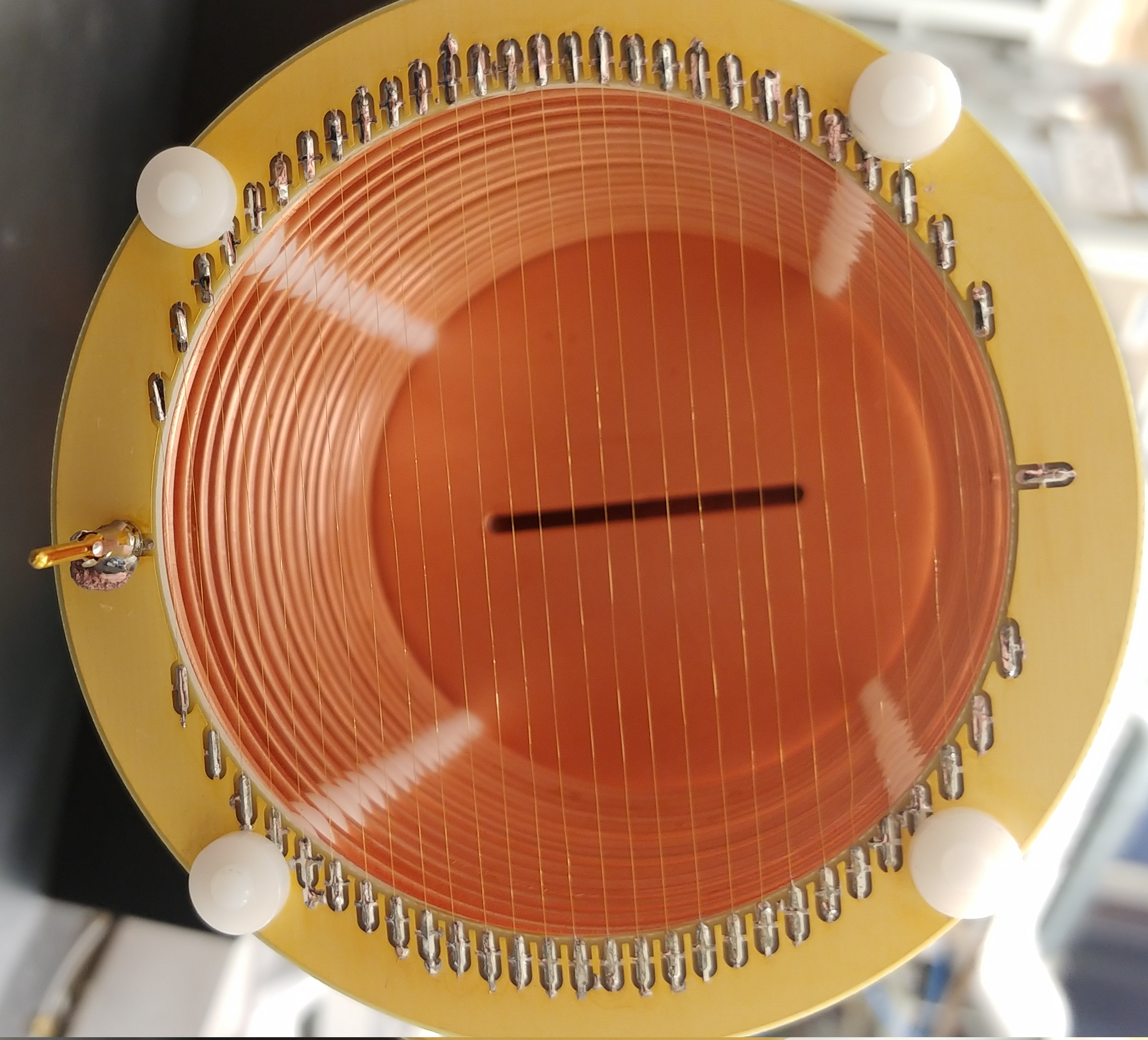}%
        };
        \begin{scope}[x={(image.south east)},y={(image.north west)}]
            \draw[blue, line width=1mm,-{latex[length=2mm, width=2mm]}](.55,.05)--(.5025,.95);%
        \end{scope}
    \end{tikzpicture}
    \caption{%
        View along the field cage, from cathode to anode plane.
        Electrons would cross vertically in a central region (blue arrow).
        The field cage is constructed from stacked copper rings and the cathode from gold-plated circuit board with crossing wires.
    }
    \label{fig:cathode_view}
\end{figure}

The \hpgmc uses a stainless steel, industry standard flanged cross pipe with modified blind flanges for electrical and gas connections to the inside (figure~\ref{fig:hpgmc_crate}) with an inner volume of $\sim\unit[4.5]{\ell}$ and a wall thickness of $\sim\unit[3]{mm}$.
In the central region, a cylindrical field cage with a single, \unit[20]{\uMicrometer} diameter, gold-plated tungsten anode wire forms the gas multiplication stage.
Figure~\ref{fig:cathode_view} shows the view along the field cage, through the wire-plane cathode.
Below the horizontal slit in the bottom plane runs the anode wire.
Electron tracks cross vertically (blue arrow), so that signals originate from tracks that cross the centre point of the field cage.

The side arms of the vessel are instrumented with the trigger system and \strontium source holders on opposite sides in exchangeable POM structures.
The front face of these structures have comb-like ridges, that fit between the field cage rings (compare figures~\ref{sfig:xsec} and~\ref{fig:sources_hv_safety}).
This yields a good precision of the alignment, especially in $\Delta z$, while avoiding a permanent fixation to the field cage or pressure vessel.
Replacement of most inner parts can be accomplished in short time and different technologies can be tested without the need for construction of a new field cage or pressure vessel.
Plastic scintillating fibres are used for the start trigger, since the trigger system only needs to provide a fast timing signal and no rejection based on calorimetric information.
By equipping the near and far fibres with two Silicon Photo Multipliers each, small signals can be separated from background (e.g. dark counts) by requiring a coincidence.

The maximum field of the \hpgmc is $\sim\unit[3000]{\uEfield}$, resulting in a reduced field of almost $\unit[90]{\uETp}$ at \unit[10]{\upbar} and room temperature.
Gas is supplied from premixed bottles and regulated with a bottle pressure regulator in high pressure operation.
A pressure relieve valve limits the over pressure of the vessel to \unit[11]{\upbar}.

The crate is equipped with pressure and temperature sensors to monitor the operation and environmental conditions.
The complete setup has been designed to be mobile for use with a host detector, e.g. a TPC at a test beam facility.

\section{Precision of Measurements}\label{sec:precision}
The maximum obtainable precision of any measurement performed with the \hpgmc can be calculated a priori by evaluating the systematic uncertainties on \estar and \vd.
Temperature and pressure measurement precisions are a combination of sensor element precision (\textit{PT100 Class A}~\cite{DINEN60751} and calibrated \textit{WIKA A10 \unit[16]{\upbar}}~\cite{WIKA-A10}) and the custom readout electronics used for these devices~\cite{Steinmann2014}.
The drift field precision is a combination of HV regulation (\textit{iSeg EPn30}~\cite{man:EPn30}), resistor chain accuracy and field cage length accuracy from measurements, see table~\ref{tab:sensor_systematics} for a list.
From~[eq.~\ref{eq:vd_definition}] follows, that uncertainties on the difference in arrival time and the track positions contribute.
The positions are determined by the minima of second order polynomial fits to the $\unit[\pm12]{\mathrm{bins}}=\unit[\pm48]{\uNanoSecond}$ region around the peak positions.
This region corresponds to the minimal width expected from the spread of arrival times caused by the \unit[1]{\uMillimeter} source's collimation.

Timing precision is given by the reconstruction accuracy of artificial drift velocity peaks (compare figure~\ref{sfig:gmc_signal}) of known $\Delta t$ with a function generator and the full analysis chain with the used digitizer (\textit{CAEN VX1720})~\cite{man:caen-vx1720,Koch:2019mpa}.
The centers of the \unit[1]{\uMillimeter} diameter $\beta$-electron beams are known to \unit[0.1]{mm}.

\begin{table}[!ht]
    \centering
    \caption{%
        Precision of observables in \hpgmc and their validity ranges.
        The reduced field precision is dominated by the used HV output precision for lower fields.
        The drift velocity is approximately equally limited by the collimation width of the used \strontium sources and time reconstruction accuracy.
    }
    \begin{tabular}{|lcc|}
        \hline
        Sensor & Range & $\sigma_{\mathrm{sys}}^{\mathrm{rel.}}$ \\
        \hline\hline
        Temperature                     & $\unit[0]{\uCelsius}<T<\unit[40]{\uCelsius}$  & \unit[$\lesssim 1$]{\permil} \\
        Pressure                        & $\unit[1]{\upbar}<p<\unit[10]{\upbar}$        & \unit[$\lesssim 2$]{\permil} \\
        \multirow{2}{*}{Electric field} & at \unit[300]{\uEfield}                       & \unit[$\lesssim 1$]{\%} \\
                                        & at \unit[3000]{\uEfield}                      & \unit[$\approx 1$]{\permil} \\
        Drift velocity                  & $\unit[5]{\uvd}<\vd$                          & \unit[$\approx 3$]{\permil} \\
        \hline
    \end{tabular}
    \label{tab:sensor_systematics}
\end{table}

\section{Drift Velocity of \pten}\label{sec:measurement}
\begin{figure}[!ht]
    \centering
    \includegraphics[width=.95\columnwidth]{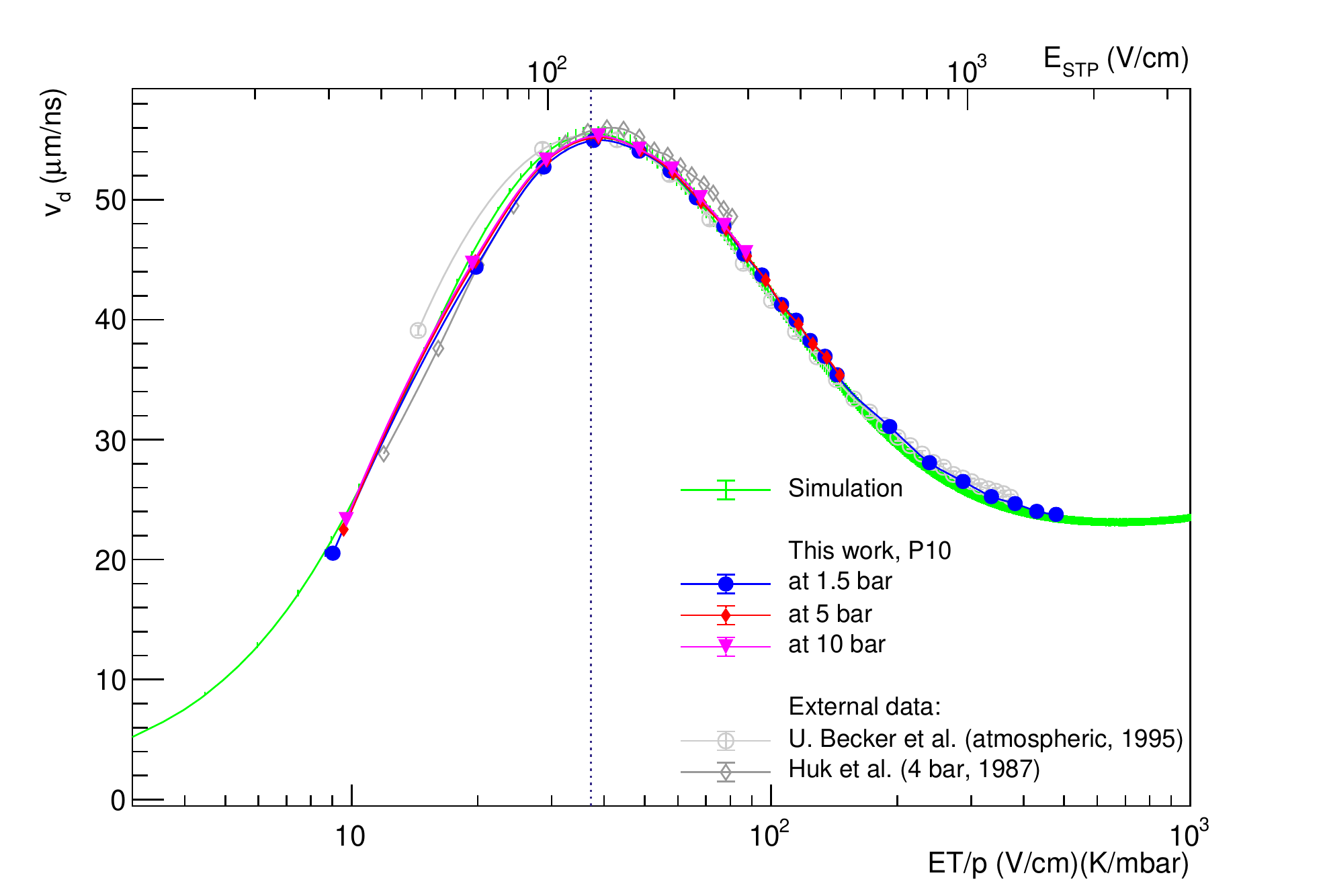}
    \caption{
        Drift velocity measurements performed by \hpgmc at three different pressure levels and external measurements~\cite{Huk:1988,Becker:1995p10}.
        The vertical line marks the location of the \vd maximum in simulation.
        Scaling with \ETp holds over the full tested span up to \unit[10]{\upbar}.
        As a reference, an additional \ESTP-axis was added, which scales \ETp to STP conditions ($p_0=\unit[1.013]{\upbar}$ and $T_0=\unit[298]{\uKelvin}$).
        Errors are statistic and systematic combined.
    }
    \label{fig:p10_hpgmc_measurement}
\end{figure}

Commissioning of the \hpgmc was carried out using premixed \pten gas (Ar:$\methane$ -- 90:10) up to a maximum pressure of \unit[10]{\upbar} in a temperature controlled laboratory at $T=\unit[(296\pm0.5)]{\uKelvin}$.
The used \pten mixture was calibrated by the distributor to consist of $\unit[10.0\pm0.2]{\umolpercent}$ \methane 4.5 in Ar 5.0.
Measurement runs were performed at three different pressure settings from $\approx\unit[1.5]{\upbar}$ up to the maximum reachable pressure of \unit[10]{\upbar} with an intermediate run at \unit[5]{\upbar}.
Figure~\ref{fig:p10_hpgmc_measurement} shows our measurements together with two external data sources~\cite{Huk:1988,Becker:1995p10} and the \magboltz expectation~\cite{Biagi:1999magb}.
Every pressure setting was maintained for a week with continuous measurements to verify a stable \vd curve.

The results show, that density scaling by [Eq.~\ref{eq:vd_etp}] holds up to \unit[10]{\upbar}.
A voltage of \unit[3600]{V} on the anode wire produced enough electron multiplication at \unit[10]{\upbar}.
This means higher pressures are only prohibitive from pressure vessel regulations and not the maximum anode voltage of \unit[5]{kV}, imposed by the used SHV feedthrough.
The amplification electronics are AC coupled to the anode wire and can decouple signals up to a \unit[10]{\ukiloVolt} DC offset.

\section{Comparison to Simulation}\label{sec:comparison}
Since the gas in a TPC is both target and medium for transport of charge signals, its drift parameters are a crucial input to detector development from the early design phase on.
Especially during the conceptional stage, researchers heavily rely on the accuracy of simulation programs for modelling.
It is therefore desirable to establish the accuracy and validity range of standard simulation programs, such as \magboltz.

\begin{figure}[!htbp]
    \centering
    \includegraphics[width=.95\columnwidth]{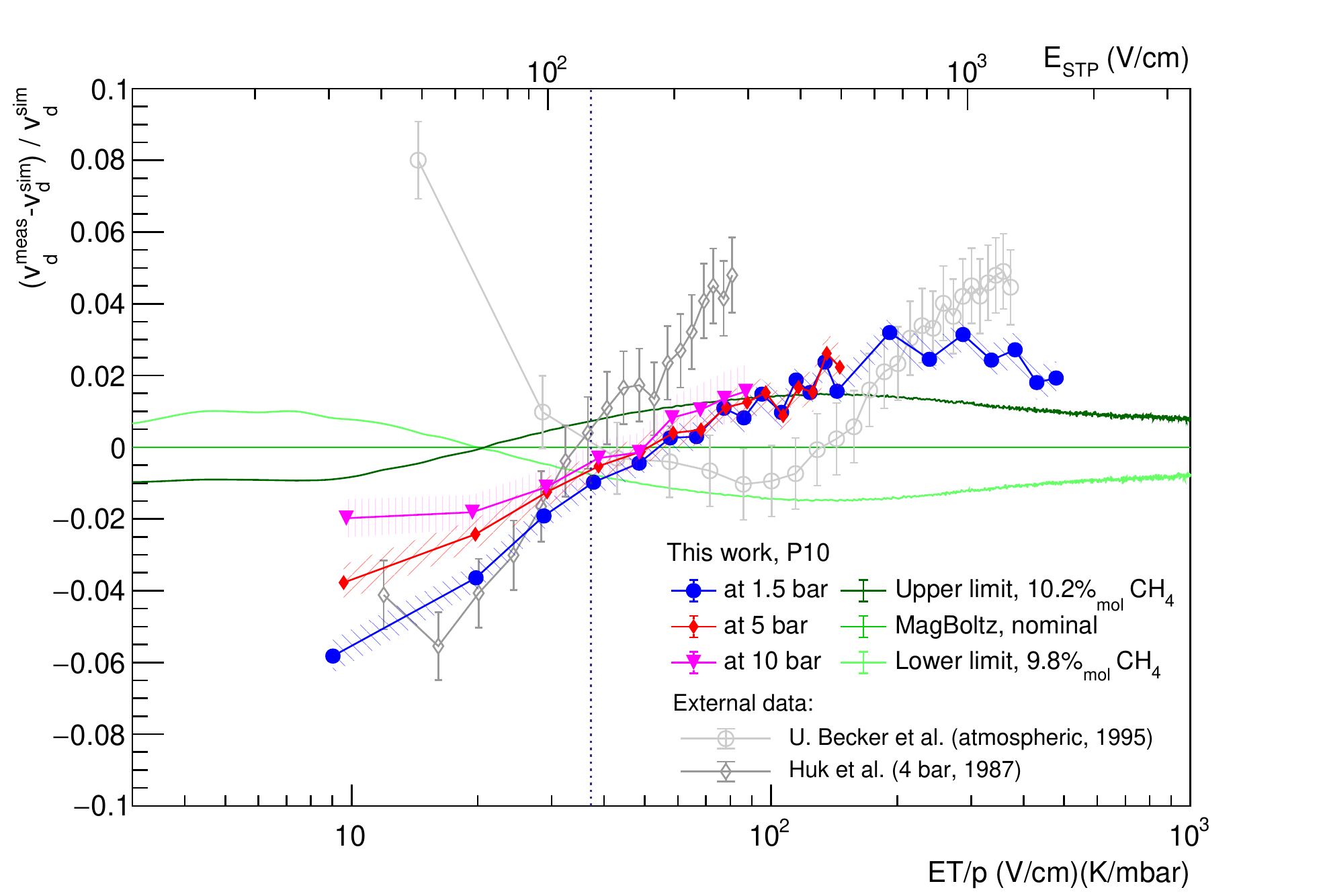}
    \caption{
        Relative deviation of measured drift velocity to simulations with \magboltz.
        The vertical line marks the location of the \vd maximum in simulation.
        Errors bands are total errors for measurements.
        Simulation errors are generally less than \unit[1]{\permil} and omitted for clearness.
    }
    \label{fig:p10_hpgmc_ratios}
\end{figure}

The relative deviation of measurements from this work and external data are shown in figure~\ref{fig:p10_hpgmc_ratios}.
Huk et al.~\cite{Huk:1988} and our measurements found slower drift velocities at fields below \unit[50]{\uETp}, compared to simulation, and faster values for fields exceeding \unit[60]{\uETp}.
The mixing uncertainty of our mixture is added as additional ratio graphs by simulating \vd with \methane content at $\unit[9.8]{\umolpercent}$ and $\unit[10.2]{\umolpercent}$.

The region of the drift velocity maximum at $\unit[\sim 38]{\uETp}$ is of particular interest for many detectors, as it offers minimal change in \vd caused by fluctuations of temperature and pressure~\cite{Abgrall:2010hi}.
At $\estar(\vd^{\mathrm{max}})$, the mixture can be expected to be sensitive to impurities due to the low total cross section caused by the Ramsauer minimum~\cite{Ramsauer:1921,BlumRolandi:2008}.
Our measurements are consistent across all pressures in that region; an indication that impurities through partial pressure driven outgassing of detector materials are negligible.
The impact of outgassing was tested by isolating the \hpgmc from the gas flow, while continuously measuring drift velocities at points around the \vd maximum and additional points at high and low \estar.
Over a duration of \unit[7]{days} and at a pressure of \unit[1.5]{\upbar}, only a \unit[-2]{\%} effect on \vd was measured at \vdmax.
At low fields, the impact of degrading gas purity was significantly larger with a reduction in \vd of almost \unit[20]{\%}.
Re-establishing the gas flow recovered the previous \vd values after $\unit[24]{h}$ and remained independent of an additionally increased gas flow.

It is notable, that there exists an \estar in simulation, for which the mixing uncertainty has vanishing impact on \vd at about \unit[20]{\uETp}.
This point lies on the rising flank of the \vd curve; a region that is particularly affected by outgassing, as observed by our tests.

\section{Summary and Conclusion}\label{sec:outlook}
The \hpgmc is a miniature TPC capable of operation with drift gases up to \unit[10]{\upbar} pressure.
It can be used for continuous calibration of a host detector or as a stand-alone experiment to study electron drift parameters.
The anode plane can be exchanged to investigate the response and stability of different multiplication geometries in novel gas mixtures at atmospheric or high pressure.
The high maximum drift field of \unit[3000]{\uEfield} covers the range of commonly employed drift fields up into the high pressure regime.
With a small inner volume of $\sim\unit[4.5]{\ell}$, rapid exchanges of the gas mixture under investigation are possible, resulting in a reduced delayed response for live monitoring.
Scaling of drift velocity curves at changing pressures was found to be consistently following $\estar=\ETp$ over a wide span of $\unit[\sim 10]{\upbar}$.
The system is highly mobile and has already accompanied a test beam experiment at CERN~\cite{deisting2021high}.
Measurements with argon based drift gases at both very low and very high, up to \unit[100]{\%}, quencher fractions are ongoing.

\begin{acknowledgments}
I would like to thank the teams in the workshops and Jochen Steinmann who made even the most complicated parts of the detector a reality and William Ma for work on the firmware of the trigger logic.
PHB is supported by DFG (Germany) Grant No. RO 3625/2-1.
\end{acknowledgments}

\bibliographystyle{apsrev4-1}
\bibliography{citations/drawings,citations/mechanics,citations/norms,citations/software,citations/electronics,citations/lkoch,citations/mixed,citations/physics_paper,citations/thesis}

\end{document}